\documentclass[journal,twoside,web]{ieeecolor}
\usepackage{generic}
\usepackage{hyperref}
\usepackage{cite}
\usepackage{adjustbox}
\usepackage{amsmath,amssymb,amsfonts}
\usepackage{algorithmic}
\usepackage{graphicx}
\usepackage{textcomp}
\usepackage{amsmath}
\def\BibTeX{{\rm B\kern-.05em{\sc i\kern-.025em b}\kern-.08em
    T\kern-.1667em\lower.7ex\hbox{E}\kern-.125emX}}
\begin{document}
\title{Single Slice Thigh CT Muscle Group Segmentation with Domain Adaptation and Self-Training }
\author{Qi Yang, Xin Yu, Ho Hin Lee, Leon Y. Cai, Kaiwen Xu, Shunxing Bao, Yuankai Huo \IEEEmembership{Senior Member, IEEE}, Ann Zenobia Moore, Sokratis Makrogiannis, Luigi Ferrucci, Bennett A. Landman \IEEEmembership{Senior Member, IEEE}
\thanks{This research is supported by NSF CAREER 1452485 and the National Institutes of Health (NIH) under award numbers R01EB017230, R01EB006136, R01NS09529, T32EB001628, 5UL1TR002243-04, 1R01MH121620-01,and T32GM007347; by ViSE/VICTR VR3029; and by the National Center for Research Resources, Grant UL1RR024975-01, and is now at the National Center for Advancing Translational Sciences, Grant 2UL1TR000445-
06. This project was also supported by the National Science Foundation under award numbers 1452485 and 2040462. This research was conducted with the support from the Intramural Research Program of the National Institute on Aging of the NIH. The content is solely the responsibility of the authors and does not necessarily represent the official views of the NIH.}
\thanks{Qi Yang, Xin Yu, Ho Hin Lee, Kaiwen Xu, Yuankai Huo are with the Computer Science Department, Vanderbilt University, Nashville TN 37215. }
\thanks{Shunxing Bao and Bennett A Landman are with the Electrical and Computer engineering Department, Vanderbilt University, Nashville TN 37215}
\thanks{Leon Cai is with the Biomedical engineering Department, Vanderbilt University, TN 37215}
\thanks{Sokratis Makrogiannis is with the PEMACS Division, Delaware State University, Dover, DE 19901}
\thanks{Ann Zenobia Moore and Luigi Ferrucci are with Translational Gerontology Branch, National Institute on Aging, NIH, Baltimore, MD 21224}}

\maketitle

\begin{abstract}
Objective: Thigh muscle group segmentation is important for assessment of muscle anatomy, metabolic disease and aging. Many efforts have been put into quantifying muscle tissues with magnetic resonance (MR) imaging including manual annotation of individual muscles. However, leveraging publicly available annotations in MR images to achieve muscle group segmentation on single slice computed tomography (CT) thigh images is challenging. 

Method: We propose an unsupervised domain adaptation pipeline with self-training to transfer labels from 3D MR to single CT slice. First, we transform the image appearance from MR to CT with CycleGAN and feed the synthesized CT images to a segmenter simultaneously. Single CT slices are divided into hard and easy cohorts based on the entropy of pseudo labels inferenced by the segmenter. After refining easy cohort pseudo labels based on anatomical assumption, self-training with easy and hard splits is applied to fine tune the segmenter.

Results: On 152 withheld single CT thigh images, the proposed pipeline achieved a mean Dice of 0.888(0.041) across all muscle groups including sartorius, hamstrings, quadriceps femoris and gracilis.
muscles

Conclusion: To our best knowledge, this is the first pipeline to achieve thigh imaging domain adaptation from MR to CT. The proposed pipeline is effective and robust in extracting muscle groups on 2D single slice CT thigh images.The container is available  for public use at \url{https://github.com/MASILab/DA_CT_muscle_seg}.

\end{abstract}


\section{Introduction}
\label{sec:introduction}
\IEEEPARstart{T}{high} muscle group segmentation is essential for assessing muscle anatomy, computing the muscle size/volume, and estimating muscle strength\cite{Hudelmaier2010}. Quantitative thigh muscle assessment from segmentation can be a potential indicator of metabolic syndrome\cite{Lim2010}. The loss of the thigh muscle and associated functional capabilities is closely related to aging\cite{Francis2017}. Accurate measurement of thigh muscle cross-sectional area, volumes and mass can help researchers understand and study the effect of aging on body composition of human body. Thus, extracting subject-specific muscle groups is an essential step.

MR imaging is the most common imaging technique in previous muscle analyses given its high contrast for soft tissue\cite{Yokota2018}. Many human efforts have been put into MR imaging for muscle analysis. Barnouin et al. optimize reproducible manual muscle segmentation\cite{Barnouin2014}. Schlaeger et al. construct a reference database (MyoSegmentTum) including the satorious, hamstring, quadriceps femoris, and gracilis muscle groups for 3D MR volume\cite{Schlaeger2018}. Compared with MR imaging, however the short acquisition time of CT is better suited for routine clinical use\cite{Yokota2018}. In longitudinal body composition study, single slice CT for each subject also reduces unnecessary radiation \cite{yu2022reducing,yang2022quantification,yu2022longitudinal}. Accurate segmentation of muscle groups on single slice can aid 
in understanding thigh components and the effects of aging on muscle\cite{overend1992thigh}.

Direct human manual annotation on single slice CT is labor-intensive and challenging due to similar intensity among different muscle groups in CT. Leveraging publicly available annotation from existing MR resources (source domain) like MyoSegmentTum for CT (target domain) is a promising direction to overcome the problem for muscle group segmentation. Methods handling domain shift or heterogeneity among modalities are called domain adaptation (DA)\cite{Guan2021}. DA aims to minimize differences among domains. DA has two challenging tasks that need to be addressed in our case: 1) homogeneous intensity of different muscle groups of CT images as mentioned before, and 2) inter-modality heterogeneity including contrast and anatomic appearance, The above two challenges can be found in Fig.1. With the above challenges for thigh muscle segmentation problems, we propose a new DA pipeline to achieve CT thigh muscle segmentation. We build a segmenter trained with synthetic CT images in CycleGAN\cite{Zhu2017}. We inference segmentation maps on real CT images by the segmenter and divide the segmentation maps into two cohorts based on entropy. The entropy can work as an indicator for prediction map quality\cite{Vu2019}. Based on anatomic context, the whole muscle and bone masks of CT images are utilized to correct wrong prediction brought by domain shift. Self-training is applied on two cohorts to make the segmenter adapt to high entropy cohorts to enhance robustness and preserve the segmentation performance on low entropy cohort

\begin{figure}
\centering
\includegraphics[width=0.4\textwidth]{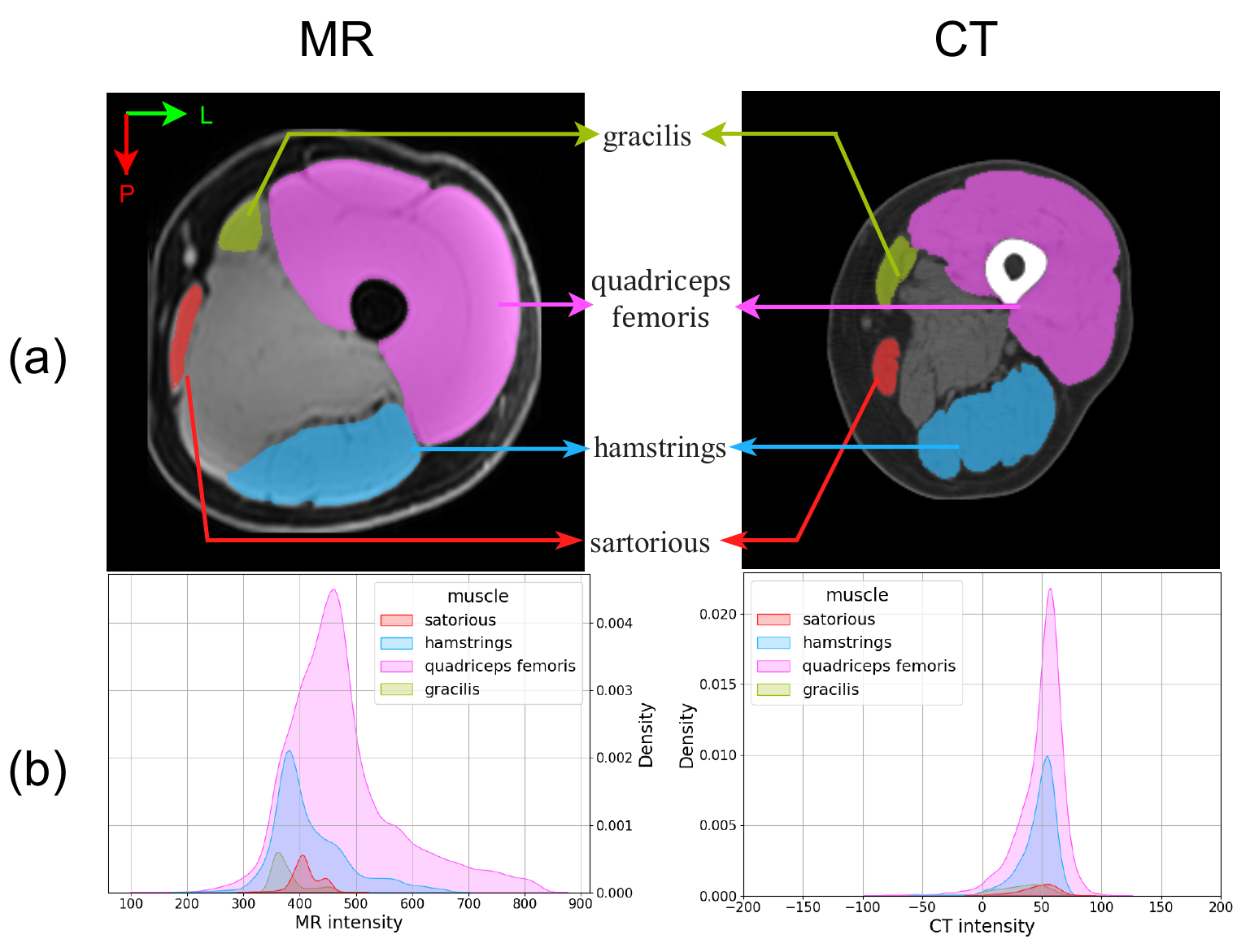}
\caption{A selective sample that highlights the inter-modality heterogeneity between MRI and CT and low intensity difference among different muscle groups in CT. (a) The MR image is normalized by min-max. The original CT scale is clipped to [-200,500] and then normalized to [0,1]. (b) is the intensity distribution for four muscle groups. The overlap intensity among four muscle groups is observed from second row.}
\label{fig:problem}
\end{figure}

Our contributions can be summarized as the following:
\begin{itemize}
\item To our best knowledge, this is the first domain adaptation pipeline for thigh muscle group segmentation on CT thigh slices. The segmenter is trained with synthetic CT images learned from the unpaired MRI dataset to have a coarse segmentation. Adversarial learning and anatomical processing of real CT images are combined to handle outliers and improve segmentation performance.
\item We evaluate the pipeline with thigh CT single slices. The experiment shows that our proposed pipeline achieves state-of-the-art performance. The ablation study demonstrates the effectiveness of anatomical processing and self-training of real CT data.
\item We release the source codes and models as singularity\cite{kurtzer2017singularity} for researchers to use and apply on their existing data.
\end{itemize}

\section{Related Work}
\subsection{Unsupervised DA}
	Unsupervised DA addresses the challenges of labeled data from source domain and unlabeled data from target domain for use during training. Unsupervised DA can reduce the amount of labor required for annotation in the target domain attracting more and more researchers, especially in medical imaging\cite{Wilson2020}.
	Dou et al. built an unsupervised DA framework for cardiac segmentation by only adapting low-level layers to reduce domain shift in training stage\cite{Dou2019}. CycleGAN\cite{Zhu2017} is known for translating image-to-image without needing pair samples, which has been applied extensively in DA. Huo et al. proposed SynSeg-Net to train CycleGAN and segmentor simultaneously to segment abdomen organs and brain image\cite{Huo2018}. Zhou et al. extended the SynSeg-Net and combined contrastive learning generative model to preserve anatomical structure during image-to-image translation\cite{Zhou2021}.  Based on image-to-image translation, Chen et al. applied a synergistic method to adapt domain invariant features and image \cite{Chen2020}. Disentanglement learning is used to learn two feature spaces: domain invariant structure and domain variant style. Yang et al. applied disentanglement learning to segment livers\cite{Yang2019} and Chang et al. embraced disentanglement learning to achieve semantic segmentation on natural images\cite{Chang2019}.

\subsection{Self-training in DA}
   
	Self-training in DA generates pseudo labels for target domain, and the model is trained by pseudo-labels data to adapt to target domain. However, directly using all pseudo labels is risky due to the accumulation of errors and domain shift negatively impacting model performance. To overcome the challenge, Zou et al. proposed to apply a re-weighting class strategy to select high-quality pseudo labels\cite{Zou2018}. Zou et al. proposed to regularize the confidence of pixel pseudo-label to adapt to target domain\cite{Zou2019}. Pan et al. proposed to reduce intra-domain gaps by dividing pseudo-labels into easy and hard splits based on entropy. The adversarial learning is applied on two splits to make model adapt to hard split without sacrificing the performance \cite{Pan2020}. To further improve the quality of pseudo-labels, anatomical prior could be incorporated into self-training procedures to correct wrong prediction caused by domain shift.

\begin{figure}
\centering
\includegraphics[width=0.4\textwidth]{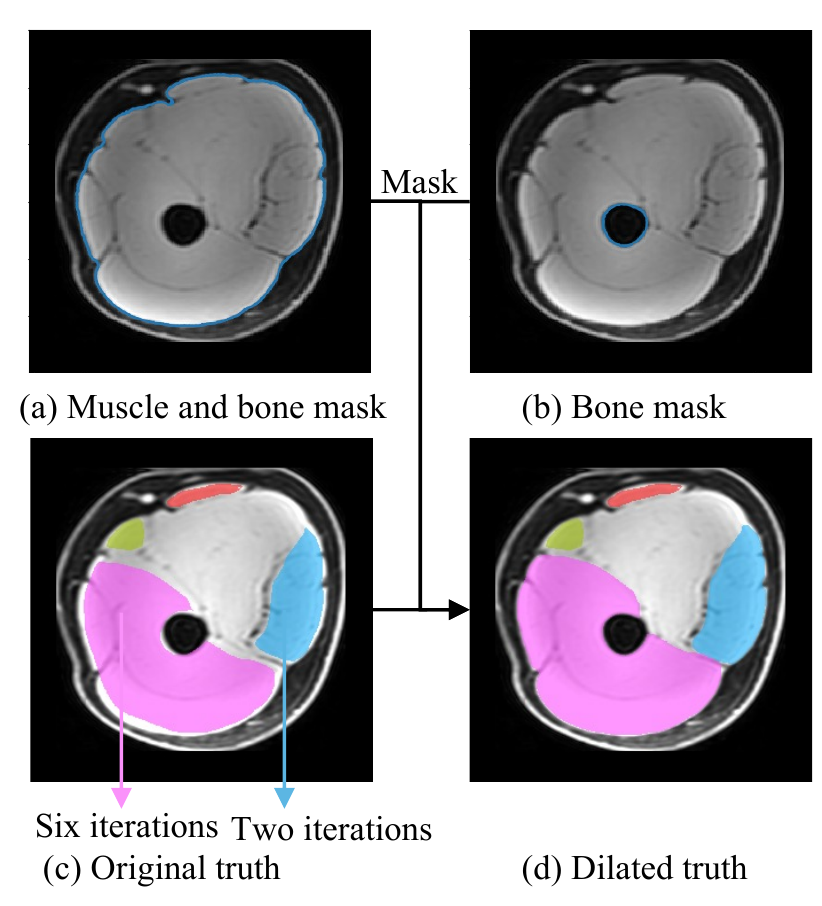}
\caption{The preprocess steps for dilating the ground truth of the MRI dataset. The blue contour in (a),(b) represent the muscle and bone boundaries extracted by level sets, and (c) represents the original ground-truth. The quadriceps femoris muscle group is dilated in 6 iterations and hamstring muscle group is dilated in 2 iterations. (d) represents the final truth after preprocessing.}
\label{fig:preprocessing}
\end{figure}

\section{Material and method}
\begin{figure*}[h!]
\centering
\includegraphics[width=\textwidth]{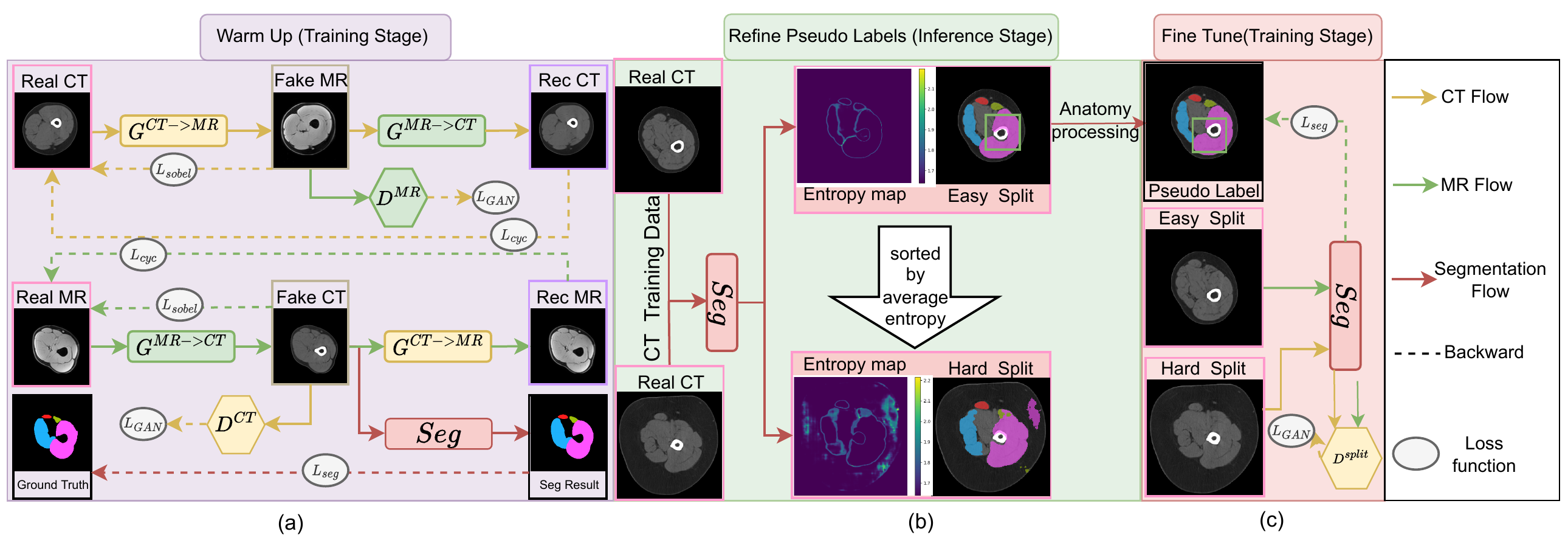}

\caption{Overview of proposed pipeline. In part (a), we adopt a CycleGAN design including two generators and two discriminators for MR and CT respectively. The segmentation module is trained by feeding synthetic CT images and corresponding MR ground truth. In part (b), the segmentation module from (a) is used to inference pseudo labels divided into hard and easy cohorts based on entropy maps. Then, the easy cohort pseudo-labels are refined based on anatomy processing (muscle and bone bask). In part (c), easy cohort pseudo-labels of CT images are used to fine tune the segmentation module, and adversarial learning between easy and hard cohorts force the segmentation module to adapt to hard cohort simultaneously to increase segmentation module robustness. }

\label{fig:method}
\end{figure*}

To solve challenge(1), (2), we proposed a pipeline includes three key parts as described in (Fig~\ref{fig:method}): (1) preprocessing on 2D single thigh slice and 3D public MRI volume, and (2) training segmentation module by feeding synthesized CT images, and (3) fine-tuning segmentation module by applying self-training on the CT training datasets.

\subsection{Data and preprocessing}
We use two datasets in our study. One is the Baltimore Longitudinal Study of Aging (BLSA)\cite{Ferrucci2008}, and the other one is MyoSegmenTUM\cite{Schlaeger2018}. The BLSA is a longitudinal dataset and collects 2D mid-thigh CT slice for each subject during the visit. BLSA study protocols are approved by the National Institutes of Health Intramural Institutional Review Board and all participants provided written informed consent. MyoSegmenTUM is a 3D MRI thigh dataset providing annotations for four muscle groups including the sartorius, hamstring, quadriceps femoris and gracilis muscle groups.

We used 1123 de-identified 2D low-dose single CT thigh slices of 763 participants from the BLSA. All data are de-identified under Institute Review Board approval. The slice has size 512×512 pixels. We split one single CT slice into left thigh and right thigh images with size 256 ×256 pixels by following the pipeline in \cite{Yang2022}. During preprocessing steps, 11 images were discarded due to low quality or abnormal anatomic appearance. The CT images are target domain in our case. 

 MyoSegmentTUM consists of water-fat MR images of 20 sessions of 15 healthy volunteers. The water protocol MR is selected as the source image. We select 1980 mid-thigh slices from MR volumes to reduce the anatomical gap between MR and CT slices at mid-thigh position. The MR slices are divided into left and right thigh images based on image morphology operation. Each image has 300 × 300 pixels.

The original label of the MR slices is placed at each group with a margin 2 mm to the outer boundary, as shown in Fig \ref{fig:preprocessing}(c). The incomplete ground truth makes the whole domain adaptation pipeline more challenging. To address this concern, we extract whole cross-sectional muscle and bone contour by using level set\cite{Li2010}. We use a binary $3 \times 3$ kernel to dilate the quadriceps femoris and hamstring muscle with six and two iterations, respectively. The complete muscle mask is obtained after performing the level set and dilation operation, as shown in Fig~\ref{fig:preprocessing}(d).

We feed random pairs of CT and MR images to the proposed method. All 1980 MR images are fed into training cohort. For CT, we divide all CT images into training, validation and test cohorts based on participants. The training cohort includes 2044 CT thigh images from 669 participants. The validation cohort consists of 38 CT thigh images from 19 participants. The test cohort consists 152 CT thigh images from 75 participants. Each CT image in the validation and test cohort have ground truth manually annotated from scratch to work for evaluation. The data distribution can be found in Table \ref{table:data}.

\begin{table*}[ht]
\caption{Data distribution and image information for whole pipeline}
\begin{adjustbox}{width=\textwidth}
\label{table:data}
\begin{tabular}{lcccc}
\hline
                     & \multicolumn{1}{l}{Participants} & \multicolumn{1}{l}{Images including left and right thigh} & \multicolumn{1}{l}{Image resolution} & \multicolumn{1}{l}{Pixel dimension (mm $\times$ mm)} \\ \hline \hline
CT training cohort   & 669                          & 2044                                                      & 256 $\times$ 256                              & 0.97 $\times$ 0.97                           \\
CT validation cohort & 19                           & 38                                                        & 256 $\times$ 256                              & 0.97 $\times$ 0.97                           \\
CT test cohort       & 75                           & 152                                                       & 256 $\times$ 256                              & 0.97 $\times$ 0.97                           \\
MR training cohort   & 15                           & 1980                                                      & 300 $\times$ 300                              & 1 $\times$ 1                                 \\ \hline
\end{tabular}
\end{adjustbox}
\end{table*}

\subsection{Train segmentation module from scratch }
Inspired by SynSeg-net\cite{Huo2018}, we design a U-Net\cite{Ronneberger2015} segmentation module ($Seg$). We train the $Seg$ with CycleGAN\cite{Zhu2017} in an end-to-end fashion as shown in Fig 2(a). CycleGAN aims to solve the image-to-image translation problem, in an unsupervised manner without requiring paired images. CycleGAN uses the idea of cycle consistency that we translate one image from one domain to the other and back again we should arrive at where we started \cite{Zhu2017}. Thus, we have two generators and discriminators in our framework. Generator $G^{X\rightarrow Y}$ represents the mapping function X:→Y. Two generators $G^{MR\rightarrow CT}$ and $G^{CT\rightarrow MR}$ are utilized to synthesis fake CT $(G^{MR\rightarrow CT}({x_{MR}}))$and fake MR $(G^{CT\rightarrow MR}(x_{CT}))$ images respectively.  The discriminator $D^{CT}$ and ${D^{MR}}$ determine whether the input image (CT or MR) is synthetic or real. The adversarial loss is applied to generators and discriminators and defined as:

\small{\begin{align}
     &L^{CT}_{GAN}(G^{MR\rightarrow CT},D^{CT},X_{MR},Y_{CT}) = 
      \mathbb{E}_{y\sim Y_{CT}}[logD^{CT}(y)] \nonumber \\ &+\mathbb{E}_{x\sim X_{MR}}[1 - logD^{CT}(G^{MR\rightarrow CT}(x))]
\end{align}}

\small{\begin{align}
     &L^{MR}_{GAN}(G^{CT\rightarrow MR},D^{MR},X_{CT},Y_{MR})= 
      \mathbb{E}_{y\sim Y_{MR}}[logD^{MR}(y)] \nonumber \\
      &+\mathbb{E}_{x\sim X_{CT}}[1-logD^{MR}(G^{CT \rightarrow MR}(x))]
\end{align}}

The above adversarial loss cannot guarantee that individual images are anatomically aligned to desired output since there are no constraints for the mapping function. Cycle loss\cite{Zhu2017} is introduced to reduce possible space for the mapping function by minimizing the difference between images and cycle reconstructed images. The loss function is:
\begin{equation}
    L_{cyc}^{CT}=||G^{MR \rightarrow CT}(G^{CT \rightarrow MR}(x_{CT}))-x_{CT}||_1
\end{equation}

\begin{equation}
    L_{cyc}^{MR}=||G^{CT \rightarrow MR}(G^{MR \rightarrow CT}(x_{MR}))-x_{MR}||_1
\end{equation}
To regularize the generator, we applied identity loss\cite{Zhu2017} to regularize generators. The identity loss is expressed as:
\begin{multline}
    L_{Identity}= 
    \mathbb{E}[||G^{MR \rightarrow CT}(x_{MR})-x_{MR}||_1] \\
    + \mathbb{E}[||G^{CT \rightarrow MR}(x_{CT})-x_{CT}||_1]
    \label{eq:identify}
\end{multline}
We further added an edge loss to preserve boundary information. Modified Sobel operator\cite{kanopoulos1988design} is utilized to extract edge magnitude. The edge loss is calculated based on the difference of edge magnitude of two images. The edge loss is expressed as Eq.\ref{eq:edge}
\begin{equation*}
     v=\begin{bmatrix}
        0 & 1 & 0 \\
        0 & 0 & 0 \\
        0 & 1 & 0
        \end{bmatrix} 
    h=\begin{bmatrix}
            0 & 0 & 0 \\
            -1 & 0 & 1 \\
            0 & 0 & 0
    \end{bmatrix}
\end{equation*}

\begin{align*}
    sobel(x,y) &=\|\sqrt{\| v * x \|_2 + \|h * x\|_2}  \\
               &- \sqrt{\| v * y \|_2 + \|h * y\|_2}\|_1
\end{align*}

\begin{align}
    L_{edge} &=sobel(G^{MR \rightarrow CT}(x_{MR}),x_{MR}) \nonumber \\ 
    &+ sobel(G^{CT \rightarrow MR}(x_{CT}),x_{CT})
\label{eq:edge}
\end{align}
where $v$ and $h$ are vertical and horizontal kernels, * represents the convolution between kernel and image. As for segmentation, weighted cross entropy loss is applied to supervise segmentation module $L_{seg}$

After defining all loss functions, we combine them together by assigning different weights $\lambda_1,\lambda_2,\lambda_3,\lambda_4,\lambda_5$ for the loss function $L$. The $L_{CT}^{GAN}$ is similar to $L_{MR}^{GAN}$ and we set the same weight $\lambda_1$ for them. $L^{CT}_{cyc}$ is symmetrical to $L^{MR}_{cyc}$, and the same weight $\lambda_2$ is assigned for those two losses. The final loss function is defined as

\begin{multline}
L=\lambda_1(L_{GAN}^{CT} + L_{GAN}^{MR}) + \lambda_2(L_{cyc}^{CT}+L_{cyc}^{MR})  \\
    +\lambda_3L_{Identity} + \lambda_4L_{edge} +\lambda_5L_{seg}
\label{eq:final}
\end{multline}

\subsection{Fine tune segmentation module in self training}
Even though we train the segmenter from scratch by feeding synthesized CT images, the segmentation module is not robust to all CT cases as shown in Fig \ref{fig:method}(b)(the segmentation map of hard split has incorrect prediction). The segmentation performance is still limited since synthetic data cannot transfer all information from real CT images.  Inspired by\cite{Pan2020}, we adopted a self-training framework to handle this challenge. We inference all pseudo labels and probability map for real CT images in training cohort. The entropy calculated based on probability for each class works as a measurement to evaluate confidence of segmentation map in unsupervised domain adaptation\cite{Vu2019}. All segmentation maps are ranked by average entropy map $I^{CT}$ from low to high. The larger the entropy, the more potential error the segmentation map includes. All segmentation maps are divided into easy and hard split based on ranking order. The first $\lambda$ of training samples are easy split and the rest is hard split. 

\begin{align}
    p^{CT} &= softmax(Seg(x^{CT})) \nonumber \\
    I^{CT} &= -\sum_{i=1}^{class}{(p^{CT}_i log_2(p^{CT}_i))}
\label{eq:entropy}
\end{align}
Where $p^{CT}$ is the probability map for each muscle class and $I^{CT}$ is the entropy map calculated based on $p^{CT}$.

Anatomical context such as spatial distribution is an important prior for medical image segmentation. To reduce incorrect prediction induced by noise and appearance shift in synthetic images, we leverage muscle and bone masks from\cite{Yang2022} to mask out erroneous predictions for easy split as show in Fig \ref{fig:method}(b).

As shown in Figure \ref{fig:method}(c), we construct the discriminator $D^{split}$ from scratch. Different from \cite{Pan2020}, the segmentation module is further trained by aligning the entropy map of easy splits to ones of hard splits. At the same time, the segmentation module is fine-tuned by feeding rectified pseudo labels of the easy split after anatomical processing and supervised by weighted cross entropy loss$L_{seg}^{easy}$. The loss function ${L^{fine tune}}$  can be expressed as:
\begin{equation*}
    L_{GAN}^{split}=\mathbb{E}_{x \sim X_{CT}^{easy}}[logD^{split}(x)] + \mathbb{E}_{y \sim Y_{CT}^{hard}}[1 - logD^{split}(y)]
\end{equation*}

\begin{equation}
    L^{fine tune}=\lambda_6L^{split}_{GAN} + L_{seg}^{easy}
\end{equation}
where ${X_{CT}^{easy}}$ is the easy split of the CT training cohort and ${X_{CT}^{hard}}$ is the hard split.${L_{seg}^{easy}}$ is a weighted cross entropy loss for the segmentation module only trained on easy cohort.

\begin{figure*}[h!]
\centering
\includegraphics[width=0.8\textwidth]{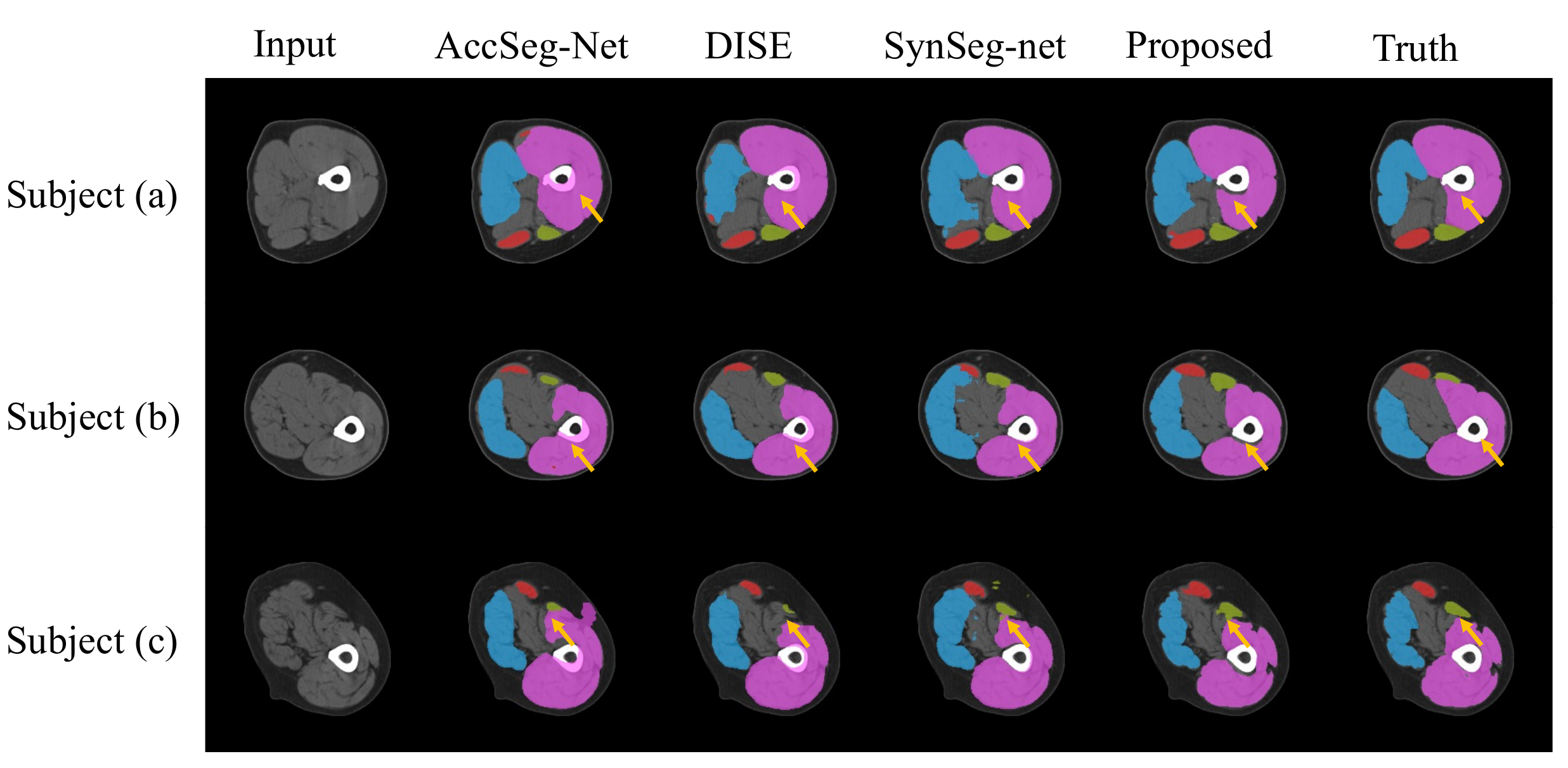}

\caption{Representation results of the proposed methods and baseline methods. Each row represents one subject.The proposed method reduces prediction errors on bones and around muscle group boundaries.The yellow arrows point to differences between proposed method and AccSeg-Net, DISE, SynSeg-net. The Input column images are rescaled for visualization purpose.}
\label{fig:qua}
\end{figure*}
    
\section{Experimental Results}
We compare the proposed pipeline with three state-of-the-art domain adaptation methods including SynSeg-net\cite{Huo2018}, AccSeg-Net\cite{Zhou2021} and DISE\cite{Chang2019}. Then we perform an ablation study to demonstrate the effectiveness of the fine-tuning stage and sensitivity analysis for proposed method.
\subsection{Implementation details and evaluation metrics}
We used python 3.7.8 and Pytorch 1.10 to implement the whole framework. The baseline and proposed methods are run on Nvidia RTX 5000 16GB GPU. For training from scratch, we set $\lambda_1$=1.0,  $\lambda_2$=30.0, $\lambda_3$=0.5, $\lambda_4$=1.0, $\lambda_5$=1.0. In the segmentation module, the weights for background, gracilis muscle, hamstring muscle, quadriceps femoris and sartorius muscle are set as [1,10,1,1,10] in the weighted cross entropy loss, respectively. For the training data divided into easy and hard cohorts, we set the first $\lambda=\frac{2}{3}$ as easy cohort and the rest as hard cohort. For fine tuning stage, we set $\lambda_6$=0.001. The initial learning rate for training model from scratch is 0.0002. We set the maximal training epochs as 100. Before first 50 epochs, the learning rate is constant as 0.0002, and then it decreases to 0 linearly. We clip original CT intensity to [-200,500].  For the MR images, we perform min-max normalization. All CT images and MR images are normalized to [-1,1]. 

Dice similarity coefficient (DSC)\cite{Dice1945} is used to evaluate the overlap between segmentation and ground truth. Briefly, we consider $S$ as the segmentation, $G$ as the ground-truth, and $||$ as the $L^1$ norm operation.
\begin{equation}
    DSC(S,G)=\frac{2|S \cap G|}{|S|+|G|}
\end{equation}
\begin{figure*}[h!]
\centering
\includegraphics[width=\textwidth]{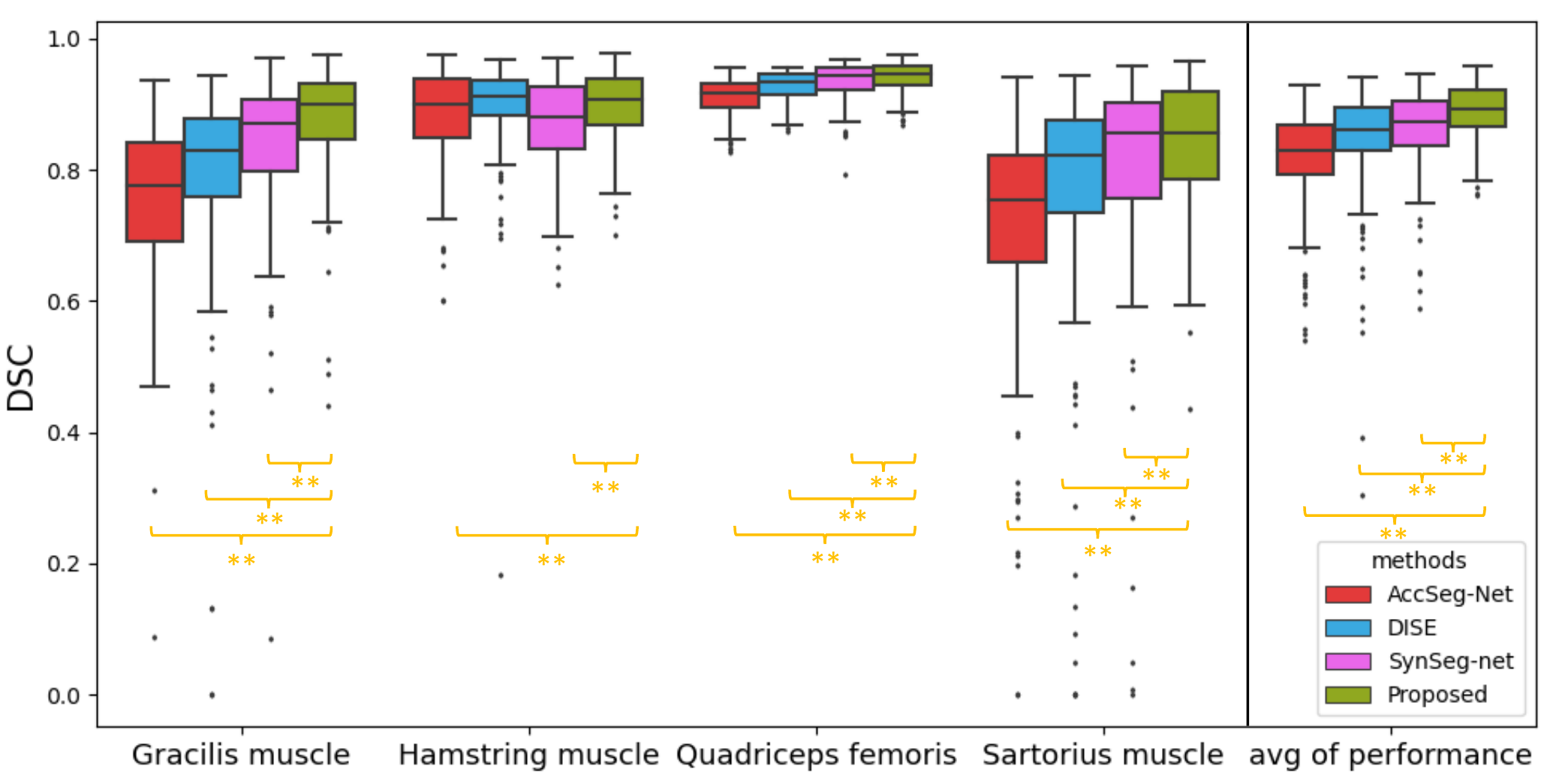}

\caption{Quantitative results of DSC of baseline methods and the proposed method. * indicates (p $<$ 0.05) significant difference between by Wilcoxon signed-rank test and ** indicates (p $<$ 0.02 corrected by Bonferroni method\cite{hollander2013nonparametric})}.
\label{fig:quan_dice}
\end{figure*}

\begin{table*}[ht]
\caption{The mean DSC and standard deviation for each muscle group and average performance. }
\begin{adjustbox}{width=\textwidth}
\label{table:quan}

\begin{tabular}{lccccc}

\hline
Method     & \multicolumn{1}{l}{Gracilis muscle} & \multicolumn{1}{l}{Hamstring muscle} & \multicolumn{1}{l}{Quadriceps femoris} & \multicolumn{1}{l}{Sartorius muscle} & \multicolumn{1}{l}{Average of four muscles} \\ \hline
AccSeg-Net & 0.753(0.128)                        & 0.882(0.075)                         & 0.91(0.028)                            & 0.708(0.176)                         & 0.813(0.08)                                 \\
DISE       & 0.786(0.159)                        & 0.895(0.078)                         & 0.928\textbf{(0.023)}                  & 0.76(0.201)                          & 0.843(0.09)                                 \\
SynSeg-net & 0.838(0.110)                        & 0.869(0.072)                         & 0.936(0.028)                           & 0.802(0.164)                         & 0.861(0.063)                                \\
Proposed   & \textbf{0.876(0.085)}               & \textbf{0.898(0.055)}                & \textbf{0.941}(0.024)                           & \textbf{0.837(0.099)}                & \textbf{0.888(0.041)}                       \\ \hline
\end{tabular}
\end{adjustbox}
\end{table*}

\subsection{Qualitative and quantitative results}
A detailed comparison of quantitative performance is shown in Table \ref{table:quan} and Fig \ref{fig:quan_dice}. All methods are trained with the same training dataset and inference is performed on the same testing dataset. From Table \ref{table:quan}, the proposed method achieves the highest mean DSC of 0.888 with lowest standard deviation of 0.041. The proposed method significantly differed from all baseline methods with $p<0.05$ under Wilcoxon signed-rank test. The proposed method achieves best DSC for each muscle group and lowest standard deviation except for the quadriceps femoris. Compared with AccSeg-Net, the proposed method makes the largest improvement in sartorious muscle increasing mean DSC from 0.708 to 0.837, and decreasing standard deviation from 0.176 to 0.099. In Fig.\ref{fig:quan_dice}, compared with the second best-performing method SynSeg-net, our method further reduces outliers and has a tighter and better DSC distribution. In Fig.\ref{fig:qua}, while the baseline methods makes incorrect prediction on bone, our method is more robust and has fewer incorrect predictions as shown in Fig.\ref{fig:qua}. 

\begin{figure}
\centering
\includegraphics[width=0.4\textwidth]{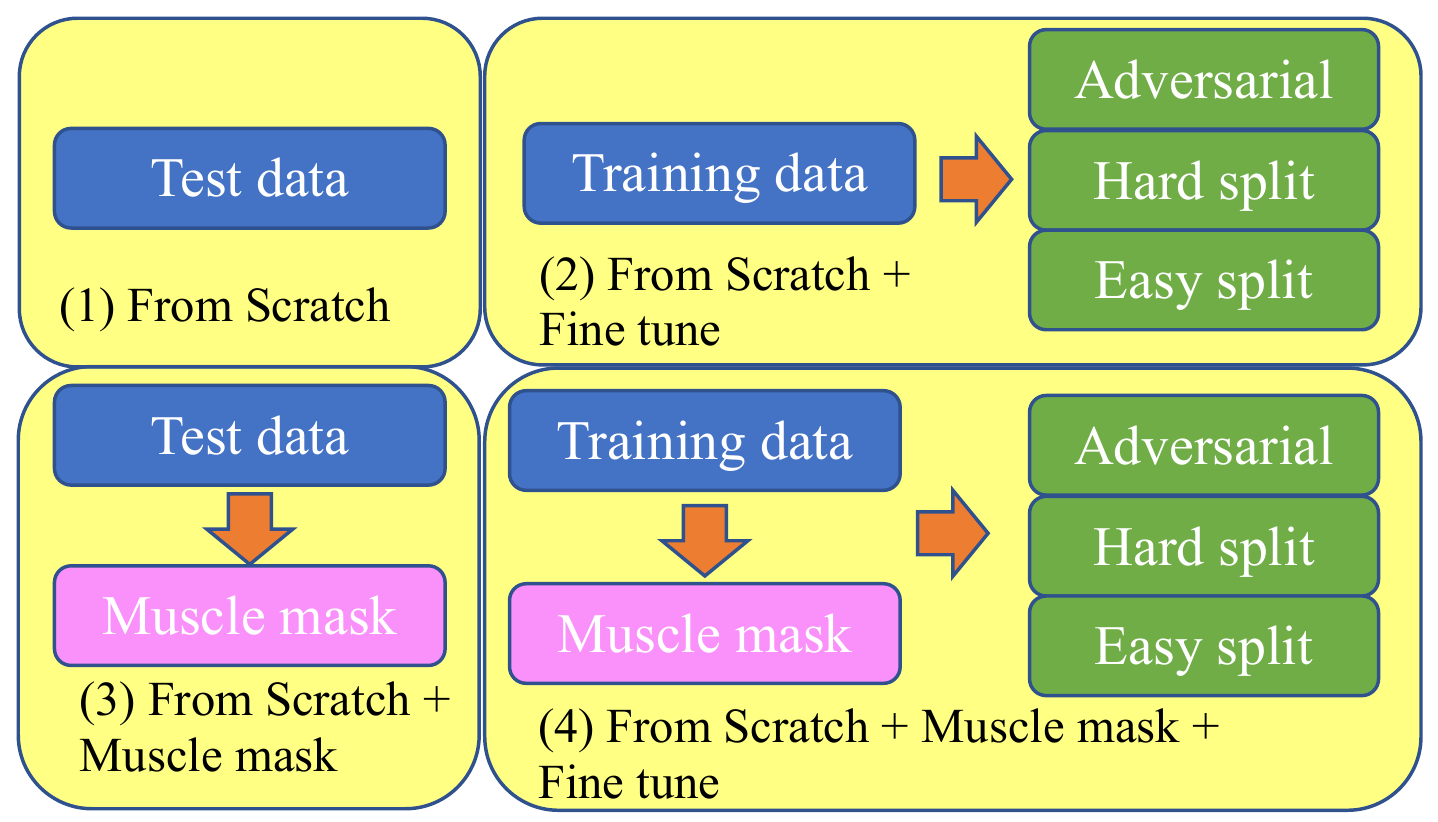}
\caption{Graphic visualization for the four pipelines designed for the ablation study. (1) represents segmentation maps inferenced by the segmentation module trained from scratch. (2) The pseudo-labels of the training data inferenced by the segmentation module from scratch and then divided in two cohorts for fine tuning. (3) The prediction map inferenced by the segmentation module from scratch is masked by muscle mask. (4) Proposed pipeline. The pseudo-labels of the training data are inferenced by segmentation module from scratch and then masked by muscle mask for fine tuning.
}
\label{fig:ablation_intro}
\end{figure}

\begin{figure*}[h!]
\centering
\includegraphics[width=\textwidth]{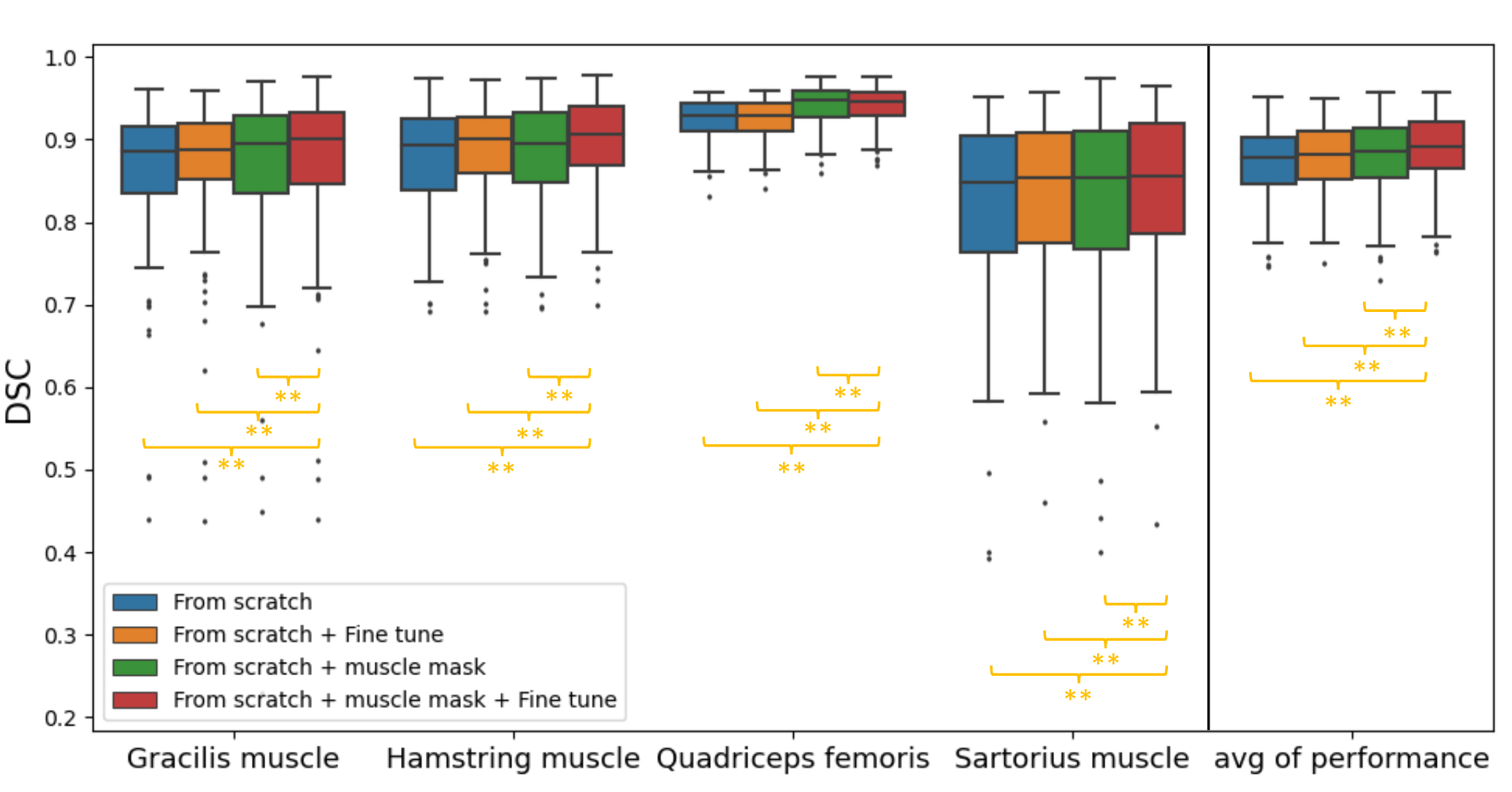}

\caption{The quantitative results for four pipelines used in the ablation study. * indicates (p $<$ 0.05) significant difference between by Wilcoxon signed-rank test and ** indicates (p $<$ 0.02 corrected by Bonferroni method.}
\label{fig:ablation}
\end{figure*}

\subsection{Ablation Study}
To investigate the effectiveness of the anatomical processing step and adversarial learning in fine tune stage, we designed 1)“From scratch”, 2)“From scratch + Fine tune”, 3)“From scratch + muscle mask” and 4)“From scratch + muscle mask + Fine tune” pipelines by modifying the procedures of the proposed pipeline. “From scratch” represents the result directly from method section B. “From scratch + Fine tune” means splitting pseudo-labels from scratch into easy and hard cohort and performing adversarial learning between two cohorts. “From scratch + muscle mask” represents that the muscle masked derived from \cite{Yang2022} is used to mask out noise for the final prediction map. “From scratch + muscle mask + Fine tune” represents the proposed pipeline. The graphic description for each pipeline is shown in Fig~\ref{fig:ablation_intro}.


As shown in Fig~\ref{fig:ablation}, compared with “From scratch”, the proposed pipeline significantly increases mean DSC from 0.870 to 0.888 and demonstrates that the anatomical processing step plus fine-tuning stage can improve segmentation performance. Compared with “From scratch + Fine tune”, the proposed pipeline significantly increased mean DSC from 0.878 to 0.888, which shows that the muscle mask can help the segmentation module discriminate noise outside the muscle mask. Compared with “From scratch + muscle mask”, the pipeline shows that adversarial learning can make segmentation module adapt to the hard split improving DSC from 0.878 to 0.888 on the test dataset instead of only relying on the muscle mask. 

\begin{figure}
\centering
\includegraphics[width=0.4\textwidth]{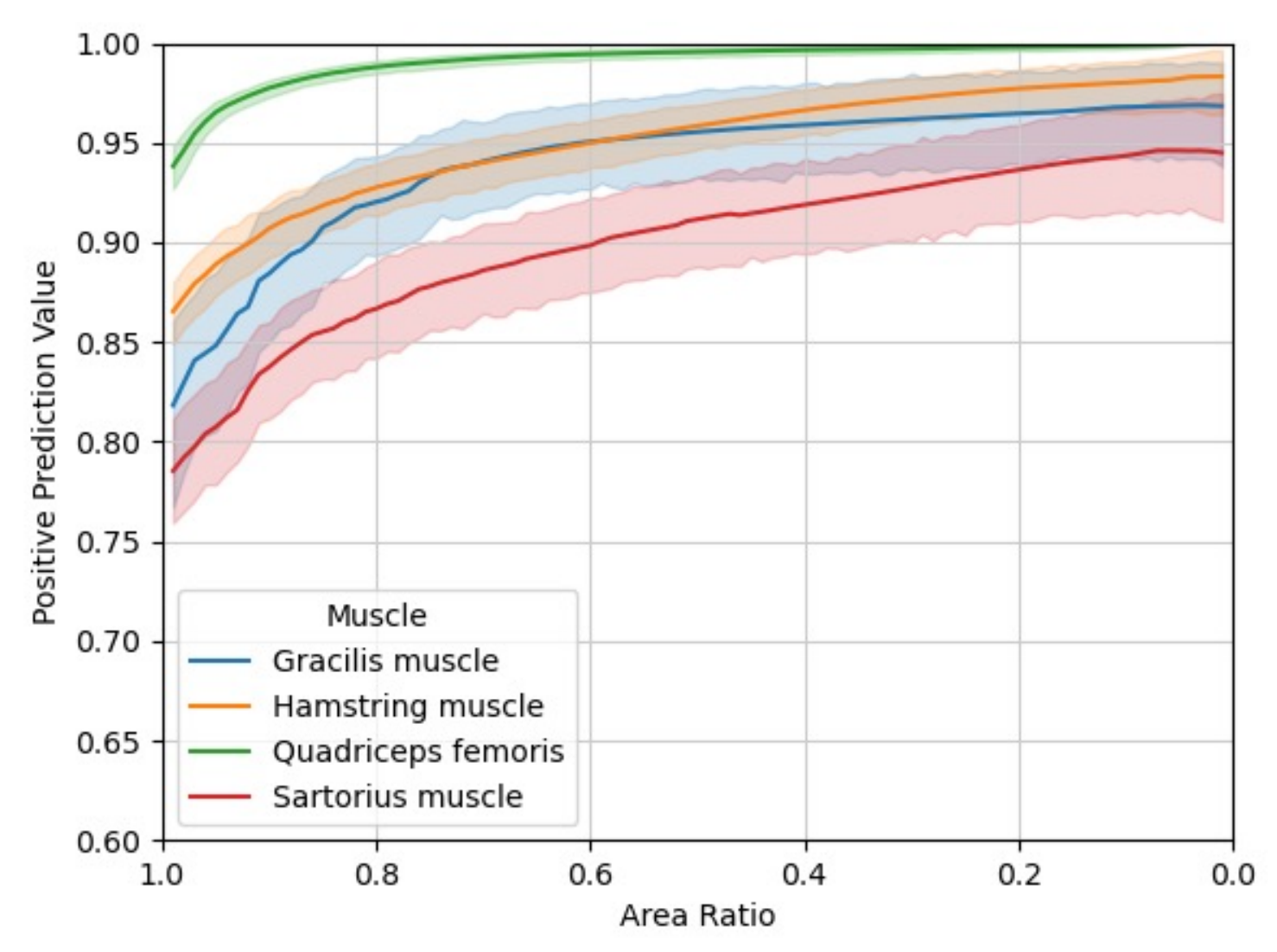}
\caption{The sensitivity plot of proposed pipeline result. The x axis represents the ratio between the eroded area and the muscle ground-truth. The positive prediction value is calculated based on Eq.~\ref{ppv}
}
\label{fig:sens}
\end{figure}

\subsection{Sensitivity analysis}
As shown in Fig.~\ref{fig:problem}, the thigh muscle is homogeneous and hard to discriminate the muscle group based on intensity alone. Furthermore, it is difficult to delineate the boundary of muscle groups by visual assessment on CT images. To check whether prediction maps cover central areas of muscle groups, we perform a sensitivity analysis to the proposed method. For each muscle group, we apply a binary 3×3 kernel to erase every muscle group iteratively until the predicted muscle group is empty. The area ratio is defined as rate between eroded muscle mask and manual ground truth. The positive predictive value (PPV) is defined as 
\begin{equation}
    PPV = \frac{|S \cap G|}{|S|}
\label{ppv}
\end{equation}
where $S$ represents the segmentation and $G$ represents the ground truth. $||$ represents the $L^1$ norm operator. From Fig \ref{fig:sens}, the quadriceps femoris has highest initial PPV of 0.94 and sartorius has lowest PPV of 0.78. The PPV of all four groups muscle are more than 0.85 when the area ratio is 0.8. The quadriceps femoris, hamstring, gracilis and sartorius muscle have a final PPV of 1.0, 0.97, 0.96, 0.95 respectively. 

\section{Discussion}

In this work, we study thigh CT and achieve single slice muscle group segmentation by proposing a two stages pipeline to leverage manual label from MR 3D volume. In the first stage, we selected single thigh CT slices from 3D volumes and split slices into left and right thigh images. The real MR and CT images were fed into a CycleGAN framework to generate synthetic CT images. The generated synthesized images were input into the segmentation module. We use the original annotation from the MR volumes to supervise the segmentation module. In the second stage, the pseudo-labels of CT images in the training cohort are inferenced by the segmentation module ($Seg$). Based on the assumption that uncertainty is related to wrong predictions, we divided the training cohort into easy and hard splits based on inference entropy. We observe that bone in MR is dark. However, bone in CT is bright. The significant contrast between MR and CT causes domain shift during CycleGAN incurring wrong prediction on the bones area. To address domain shift problem, the muscle mask from \cite{Yang2022} is used to correct noise map. Finally, adversarial learning is utilized to align the prediction map between easy and hard split to make segmentation module robust to real CT images. To our best knowledge, this is the first pipeline to perform domain adaptation on thigh CT images. We collect all modules into one container to let the public and more researchers take advantage of our contribution. The segmentation module can be directly used for single slice CT muscle group segmentation.

Although the proposed pipeline can handle current challenges in domain adaption, limitations still exist in the process of the proposed pipeline. One limitation is the dependence on pseudo labels when training from scratch. need researchers to empirically tune the The hyperparameters need to tuned empirically to make the generative model to synthesis anatomy consistent images since we need to balance the generator and discriminator simultaneously. Another limitation is that even though the entropy map is closely related with noise, prediction errors cannot be found only based on entropy maps. It means that the segmentation module might learn incorrect patterns in the fine tuning stage and needs to further study, which is beyond the scope of this manuscript.

\section{Conclusion}
In summary, we present a novel pipeline to leverage muscle group annotations from MR 3D volumes in segmenting single thigh CT slices. In this study, we (1) proposed a pipeline to solve the domain adaptation problem for CT thigh images, (2) applied the proposed pipeline on CT thigh images and demonstrated effectiveness and robustness of pipeline, and (3) packed all modules into a container for researchers to extract muscle groups conveniently and directly without manual annotation. As our current pipeline includes multi-stages, the way to improve the whole pipeline is to bundle into one end-to-end framework.
\bibliographystyle{IEEEtran}
\bibliography{refs}
\end{document}